# Detection of Beat-to-Beat Intervals from Wrist Photoplethysmography in Patients with Sinus Rhythm and Atrial Fibrillation after Surgery

Adrian Tarniceriu, Jarkko Harju, Antti Vehkaoja, Jakub Parak, *IEEE Student Member*, Ricard Delgado-Gonzalo, Philippe Renevey, Arvi Yli-Hankala, and Ilkka Korhonen, *IEEE Senior Member*

*Abstract—* Wrist photoplethysmography (PPG) allows unobtrusive monitoring of the heart rate (HR). PPG is affected by the capillary blood perfusion and the pumping function of the heart, which generally deteriorate with age and due to presence of cardiac arrhythmia. The performance of wrist PPG in monitoring beat-to-beat HR in older patients with arrhythmia has not been reported earlier. We monitored PPG from wrist in 18 patients recovering from surgery in the post anesthesia care unit, and evaluated the inter-beat interval (IBI) detection accuracy against ECG based R-to-R intervals (RRI). Nine subjects had sinus rhythm (SR, 68.0y ± 10.2y, 6 males) and nine subjects had atrial fibrillation (AF, 71.3y ± 7.8y, 4 males) during the recording. For the SR group, 99.44% of the beats were correctly identified, 2.39% extra beats were detected, and the mean absolute error (MAE) was 7.34 ms. For the AF group, 97.49% of the heartbeats were correctly identified, 2.26% extra beats were detected, and the MAE was 14.31 ms. IBI from the PPG were hence in close agreement with the ECG reference in both groups. The results suggest that wrist PPG provides a comfortable alternative to ECG and can be used for long-term monitoring and screening of AF episodes.

## I. Introduction

Heart rate variability (HRV) provides significant information about a person's health status. It is used for sleep analysis [1], stress and recovery analysis [2], and also in clinical applications such as atrial fibrillation (AF) detection [3, 4, 5].

Traditionally, ECG devices have been used in data collection for HRV analysis. The most common are chest straps and electrode patches, which can provide high accuracy for the estimation of beat-to-beat intervals [6, 7, 8] when compared to ambulatory ECG recorders, but can become uncomfortable when being worn for longer durations. In addition, dry skin or poor skin contact often disturb chest strap based HRV monitoring. Thus, there is a clear demand for new technologies, which do not interfere with a person's comfort.

Photoplethysmography (PPG) provides an alternative method for HR and HRV monitoring [9]. The skin is illuminated with a LED and a photodetector measures the intensity of the transmitted or reflected light. This intensity depends on the blood volume in the skin capillaries and the vasculature deeper in the tissue, which, in turn, vary with the pumping actions of the heart. Thus, by analyzing the light intensity, we can determine the heart rate and inter-beat intervals (IBI).

Currently, optical heart rate (OHR) devices can provide adequate accuracy for heart rate estimation during rest, sports, and daily activities [10, 11]. Previous work [12, 13] showed that using the right algorithms, IBI can be estimated from wrist PPG signals with errors below 10 ms, which is accurate enough for HRV analysis. However, these results were obtained using data from healthy working age subjects. Elderly people usually have poorer peripheral perfusion, different skin structure, and arrhythmias or other illnesses. All these factors affect the PPG signal, decreasing the signal-to-noise ratio.

This study evaluates the IBI estimation accuracy for a group of post-surgery patients, half of which suffer from AF. The main goal is to evaluate whether wrist PPG can be used for IBI monitoring in clinical applications for elderly subjects with arrhythmia and especially with AF. If proven operational, the wrist PPG technology would provide tremendous benefits in both clinical and home monitoring scenarios: it would provide a comfortable, wearable, unobtrusive measurement method suitable for long-term monitoring. Besides life-style, sleep, and stress analysis, it could be also used in screening of various cardiac anomalies.

## II. Materials and Methods

### A. Subjects

All recordings took place in the post-anesthesia care unit of the Tampere University Hospital. The patients had undergone surgery immediately prior to the recording and were recovering from the effects of anesthetics. They were laying down in bed during the whole duration of the measurement. The average duration of each recording is 1.5 hours. 18 patients were included and classified into two groups: with sinus rhythm (SR) and with continuous AF during the recording. The SR group consisted of nine subjects - six male, three femane, 68.0 ± 10.2 years old, and the AF group consisted of nine subjects - four male, five female, 71.3 ± 7.8 years old.

A. Tarniceriu is with PulseOn SA, Jacquet-Droz 1, 2002, Neuchâtel, Switzerland (corresponding e-mail: adrian.tarniceriu@pulseon.com).

J. Harju and A. Yli-Hankala are with the Tampere University Hospital, Tampere, Finland.

A. Vehkaoja, J. Parak, and I. Korhonen are with BioMediTech Institute and the Faculty of Biomedical Sciences and Engineering, Tampere University of Technology, Tampere, Finland, and with PulseOn Oy, Espoo, Finland.

R. Delgado-Gonzalo and P. Renevey are with CSEM - Centre Suisse d'Electronique et Microtechnique, Jacquet-Droz 1, 2002 Neuchâtel, Switzerland.

The study protocol, devices, and documentation were approved by the local ethical review board of Pirkanmaa Hospital District (R17024), the Finnish National Supervisory Authority of Health and Welfare, and the technical department of the hospital. The test subjects gave their written consent to participate after being informed on the purpose of the study and they had the right to withdraw from the study at any time. The experimental procedures comply with the principles of the Helsinki Declaration of 1975, as revised in 2000.

*B. Data Acquisition*

Wrist PPG signals were recorded with the PulseOn OHR tracker (www.pulseon.fi), presented in Fig. 1. The device was worn as instructed by the manufacturer, about one finger width from the wrist bone and tightened by the person in charge of data collection so that the skin contact was firm but still comfortable for the whole recording. For the PPG data, the IBI were provided by OHR tracker directly.

The ECG signals were measured with the GE Healthcare Carescape B850 (www.gehealthcare.com) patient monitor and recorded with the S5 Collect software. The RR intervals were obtained from the ECG signal using the Kubios HRV software, version 2.2 (www.kubios.com). The ECG waveforms were also visually inspected to ensure that no R-waves are missed.

*C. Methods*

As the recording of wrist PPG and ECG signals did not start at the same time, we firstly synchronized the IBI and RRI time series. This was done by compensating for eventual time drifts between the PulseOn and Carescape B850 clocks and by minimizing the mean absolute error between the IBI and reference RRI vectors. For final synchronization, we split the data in intervals of one minute and performed a new synchronization for each interval. This was necessary to allow beat-to-beat level synchronization despite slightly differing nominal clock rates of the devices. Ectopic beats [14] were excluded from the evaluation.

In the next step, for each one-minute interval, we determined the percentage of correctly detected beats, extra beats, and missed beats with respect to the ECG reference. This was done with a method similar to the one used in [13]. For every PPG-detected beat, we checked how many reference beats were detected in the interval $[t - 0.5l, t + 0.5l]$, where $t$ is the time when the beat was detected and $l$ is the length of the corresponding IBI. If there was only one reference beat, then it was correctly detected. If there were no corresponding reference beats, then an extra beat had been detected. The reference beats with no corresponding PPG-detected beat were considered missing beats.

Most extra and missing beats are explained by the fact that IBI estimation is not accurate during motion. An example is given in Fig. 2: motion, depicted as variations in the 3D acceleration signal, causes more oscillations in the PPG-based IBI signal. This reduces the accuracy of beat estimation [15], usually resulting in shorter IBI, as seen between 15 and 35s. These type of artefacts can occur even if the movement is limited to the fingers or hand, and the forearm is immobile, movements that are not detected by an accelerometer located in the wrist device.

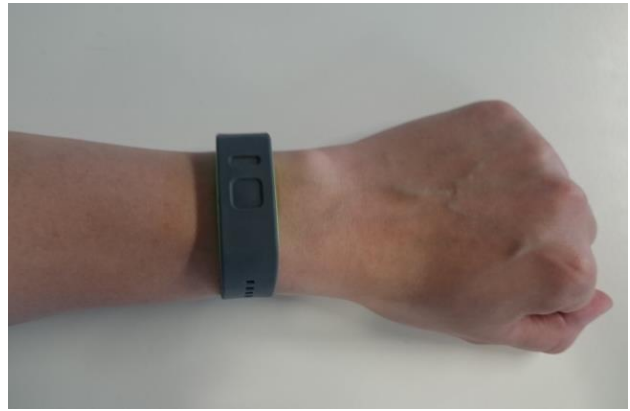

Figure 1. PulseOn OHR tracker placed on the wrist

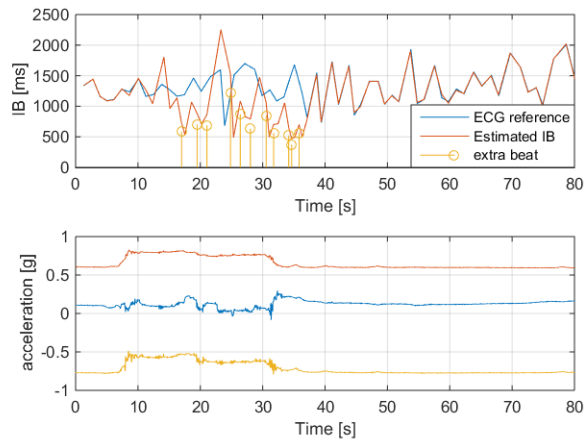

Figure 2. Effect of motion, depicted as variations in the 3D acceleration signal, on the estimation of inter-beats from wrist PPG signals.

In addition to extra detected and missed beats, we compute the mean error (ME), the mean absolute error (MAE), the mean absolute percentage error (MAPE), and the root mean square error (RMSE) for the IBI-RRI pairs. Three HRV parameters, were also computed: the root mean square of successive differences (RMSSD), the percentage of successive IBI that differ by more than 50 ms (pNN50), and the IBI standard deviation (STD), to evaluate their behavior for SR and AF rhythms. As the purpose of this study is to estimate the beat accuracy during rest, and the missing or extra beats are a good indicator for the presence of motion, we will only consider the one-minute intervals with no missing or extra beats when computing the ME, MAE, MAPE, RMSE, RMSSD, pNN50, and STD.

### III. RESULTS AND DISCUSSION

*A. Beat Detection Performance*

The beat detection results are summarized in Table I. For the SR set, 99.44% of the beats were correctly detected while for the AF set, 97.49% of the beats were correctly detected. The level of extra beats is similar between the groups (2.39% vs 2.26% for SR and AF, correspondingly), while the AF group has significantly more missing beats than the SR group (2.51% vs 0.56%). The lower beat detection rate in the AF group can be explained by different pulse morphology caused by arrhythmias.

TABLE I. IBI DETECTION PERFORMANCE

|  | SR set | AF set |
|---|---|---|
| Total beats | 52726 | 55565 |
| Correct beats [%] | 99.44 | 97.49 |
| Extra beats [%] | 2.39 | 2.26 |
| Missing beats [%] | 0.56 | 2.51 |

TABLE II. IBI ESTIMATION PERFORMANCE

|  | SR set | AF set |
|---|---|---|
| ME [ms] | -0.40 | -0.47 |
| MAE [ms] | 7.34 | 14.31 |
| MAPE [%] | 0.79 | 1.58 |
| RMSE [ms] | 16.70 | 23.52 |

*B. IBI Estimation*

Figures 3 and 4 illustrate an example of 50 beats extracted from the PPG signals as well as from the ECG reference, and the error between IBI and RRI for SR and AF cases, respectively. The difference between SR and AF scenarios is clearly visible from these figures, the AF case having a much higher variation between consecutive IB values.

The MAE and MAPE are approximately two-fold higher for the AF group as compared to the SR group (Table II). Fig. 5 shows the Bland-Altman plots for the RRI and IBI. The most likely explanation for the higher error in the AF group is that the fiducial point of the pulse wave detection in PPG is dependent on the pulse morphology, which is widely variant during AF due to non-optimal heart filling and poor pumping function. However, the MAE for the AF group is still significantly lower than the difference between the consecutive beats, as can be seen in Fig. 4, and each case follows the general trend of the RRI values extracted from the ECG signals.

The estimation error is slightly biased towards lower values (the ME is -0.40 and -0.47 ms, respectively), most likely due to the rounding towards zero operations of the used fixed-point algorithm. This error, lower than 1 ms, has a negligible effect on HRV analysis.

For the AF group, there is no visible correlation in the Bland-Altman plot between the IBI-RRI difference and the values of the IBI. For the SR group, it looks like the error dispersion is higher for beats of ~1000 ms. However, this is just a visual effect of the fact that there are more beats around this value. For 7 out of 9 sets, the average HR is between 55 and 65 beats per minute, corresponding to IBI between 923 and 1090 ms; but the error standard deviation is the same for beats of ~1000 ms and for beats with different durations.

*C. Heart Rate Variability Parameter Comparison*

Table III presents three HRV parameters calculated from IBI in SR and AF groups. The HRV parameters are systematically higher for the AF group, suggesting that they may be used to differentiate AF from SR [17, 18]. Another insight on the usability of PPG-derived inter-beats for the detection of atrial fibrillation is provided in Fig. 6. Here, we plot the standard deviation of groups of 20 consecutive IB values. The difference between SR and AF cases is clearly visible, and one could easily distinguish between the two cardiac conditions. This can be use as the starting point for designing an atrial fibrillation detection algorithm.

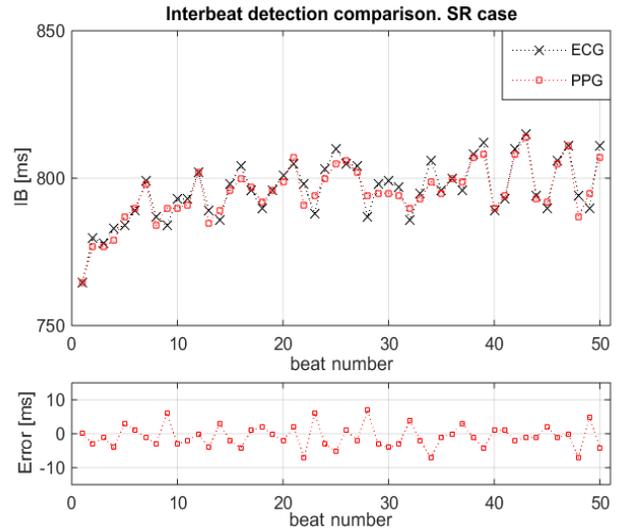

Figure 3. Example of IBI and RRI time series in a SR case. The lower panel shows the instantaneous error between RRI and IBI

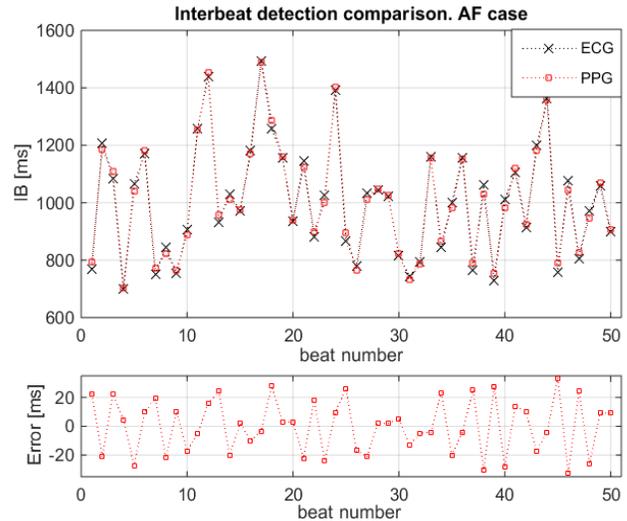

Figure 4. Example of IBI and RRI time series in an AF case. The lower panel shows the instantaneous error between RRI and IBI

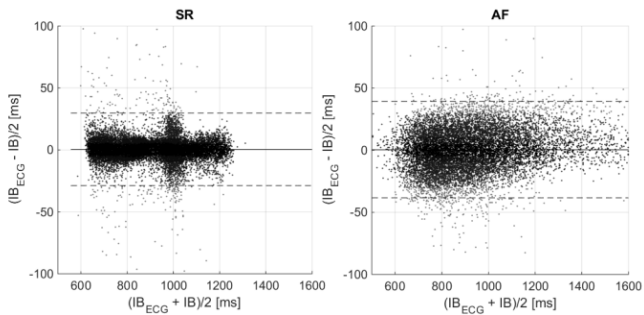

Figure 5. Bland-Altman plots for PPG IB intervals, relative to the ECG reference. Sinus rhythm and atrial fibrillation cases

TABLE III. HRV STATISTICS

|  | SR set | AF set |
|---|---|---|
| RMSSD [ms] | 36.01 | 268.34 |
| pNN50 [%] | 8.45 | 83.09 |
| STD [ms] | 51.70 | 211.48 |

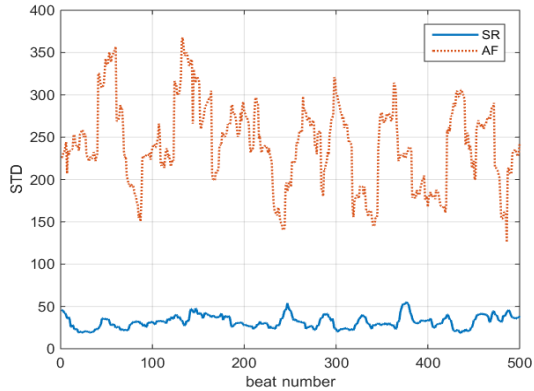

Figure 6. Standard deviation example for groups of 20 consecutive inter-beat values for AF and SR cases

## IV. CONCLUSION

This study evaluated the accuracy of IBI estimation from wrist PPG signals for elderly patients after surgery with SR and AF. The MAE values are 7.34 ms for the SR group and 14.31 ms for the AF group. This is accurate enough for both HRV analysis and to differentiate between SR and AF cases.

Earlier studies have validated the estimation of IBI from PPG signals for healthy subjects during sleep [13]. This study validates the IBI estimation in a more challenging scenario: the subjects are elderly patients with arrhythmia, and have undergone surgery prior to the recording. For comparison, the MAE value obtained in [13] is 6.68 ms which is almost identical to MAE observed in this study for SR patients.

In conclusion, the present study confirms that IBI estimated from wrist PPG signals are in close agreement with RRI obtained from the ECG reference. The estimated values are highly accurate and can be used for both HRV analysis and clinical applications such as AF detection. This provides a promising alternative to current monitoring technologies, and an important step towards 24/7 monitoring.